\documentclass{elsart}
\usepackage{graphicx}
\usepackage{epstopdf}

\def\kslash{k\kern-.5em\slash}
\def\Pslash{P\kern-.5em\slash}
\def\pslash{p\kern-.5em\slash}
\def\qslash{q\kern-.5em\slash}
\def\dslash{\partial\kern-.5em\slash}

\begin{document}
\begin{frontmatter}
\title{Effective interactions from $q$-deformed inspired transformations}
\author[CESET]{V. S. Tim\'oteo}
\author[IFUSP]{C. L. Lima}
\address[CESET]{Centro Superior de Educa\c c\~ao Tecnol\'ogica,
Universidade Estadual de Campinas,
13484-370, Limeira, SP, Brasil}
\address[IFUSP]{Instituto de F\'{\i}sica, Universidade de S\~{a}o Paulo,
CP 66318, 05315-970, S\~{a}o Paulo, SP, Brazil}
\date{\today}
\maketitle
\begin{abstract}
From the mass term for the transformed quark fields, we obtain effective contact interactions of the NJL type. 
The parameters of the model that maps a system of non-interacting transformed fields into quarks interacting 
via NJL contact terms are discussed.
\end{abstract}
\end{frontmatter}

It is very common in physics to use transformations that make one particular
system mathematically simpler, yet describing the same phenomena. A clear
example is the use of canonical transformations in classical mechanics.

$q$-Deformed algebras provide a nice framework to incorporate, in an effective
way, interactions not originally contained in the Lagrangian of a particular
system. 

In hadron physics, the NJL model is a very simple effective model for strong
interactions that describes important features like the dynamical mass
generation, spontaneous chiral symmetry breaking, and chiral symmetry
restoration at finite temperature.

In recent works, we have been investigating possible applications of quantum
algebras in hadronic physics. In general, we observed that when we deform the
underlying algebra, the system is affected with correlations between its
constituents. We have studied in detail the NJL model under the influence of a
quantum $su(2)$ algebra.

The question we approach in this letter is: is it possible to obtain a transformation 
connecting the NJL model to a simpler
non-interacting system? We verified that we can indeed obtain the same dynamics 
of the NJL interaction with a simple transformation of the quark fields, inspired 
in the $q$-deformed quark fields of previous works \cite{ubri}, \cite{trip}, \cite{VST}.

\section*{Mass term}

We start by writing a mass term for the transformed quark fields
\begin{eqnarray}
\mathcal{L}_{q}^{mass}  &=& -M~\overline{\Psi}\Psi\nonumber\\
&=& -M~\left(  ~\overline{\Psi}_{1}\Psi_{1}+\overline{\Psi}_{2}\Psi
_{2}~\right) \nonumber \\
&=& -M~\left(  ~\overline{U}U+\overline{D}D~\right) 
\label{Lq}
\end{eqnarray}
where
\begin{equation}
\Psi=\left(
\begin{array}
[c]{cc}
\Psi_{1} & \\
\Psi_{2} &
\end{array}
\right)  =\left(
\begin{array}
[c]{cc}%
U & \\
D &
\end{array}
\right)  ~.
\end{equation}

The transformed quark fields can be written in terms of the standard fields as
\begin{eqnarray}
\Psi_{1}  &=& \psi_{1} + 
(q^{-1} - 1)~\psi_{1}\overline{\psi}_{2}\gamma_{0}
\psi_{2}~,  \\
\Psi_{2}  &=& \psi_{2} + 
(q^{-1}-1)~\psi_{2}\overline{\psi}_{1}\gamma_{0}
\psi_{1}~, 
\label{dff}
\end{eqnarray}
or
\begin{eqnarray}
U  &=& u+(q^{-1}-1)~u\overline{d}\gamma_{0}d~,\label{duf} \\
D  &=& d+(q^{-1}-1)~d\overline{u}\gamma_{0}u~, \label{ddf} 
\end{eqnarray}
and 
\begin{equation}
\psi=\left(
\begin{array}
[c]{cc}
\psi_{1} & \\
\psi_{2} &
\end{array}
\right)  =\left(
\begin{array}
[c]{cc}
u & \\
d &
\end{array}
\right)  ~.
\end{equation}
Here both components are modified in the same way, so that the above 
expressions are different from the ones used in \cite{ubri,trip}, where only one
component is affected. Extending the transformation to both components
is required to obtain a set of terms that will form an interaction of the NJL 
type. This implies that the anti-commutation relations for the deformed fields
$\Psi$ will also be different from the ones in \cite{ubri,trip}. Since obtaining the new
anti-commuation relations is not in the scope of this work, we focus on the effective
interactions contained in the non-interacting Lagrangian.

Using Eqs. (\ref{duf}) and (\ref{ddf}), we can re-write the Lagragian
Eq.(\ref{Lq}) in terms of the standard quark fields
\begin{eqnarray}
\overline U U  &=& \overline u u + Q~
\overline u u d^{\dagger}d + Q~d^{\dagger}d \overline u u + Q^{2}~ \overline d d \overline u u \overline d d~ , \\
\overline D D  &=& \overline d d + Q~
\overline d d u^{\dagger}u +Q~u^{\dagger}u \overline d d + Q^{2}~ \overline u u \overline d d \overline u u~ ,
\end{eqnarray}
where $Q=(q^{-1}-1)$.

\begin{figure}[t]
\centering           
\includegraphics[width=6cm ,height=5cm]{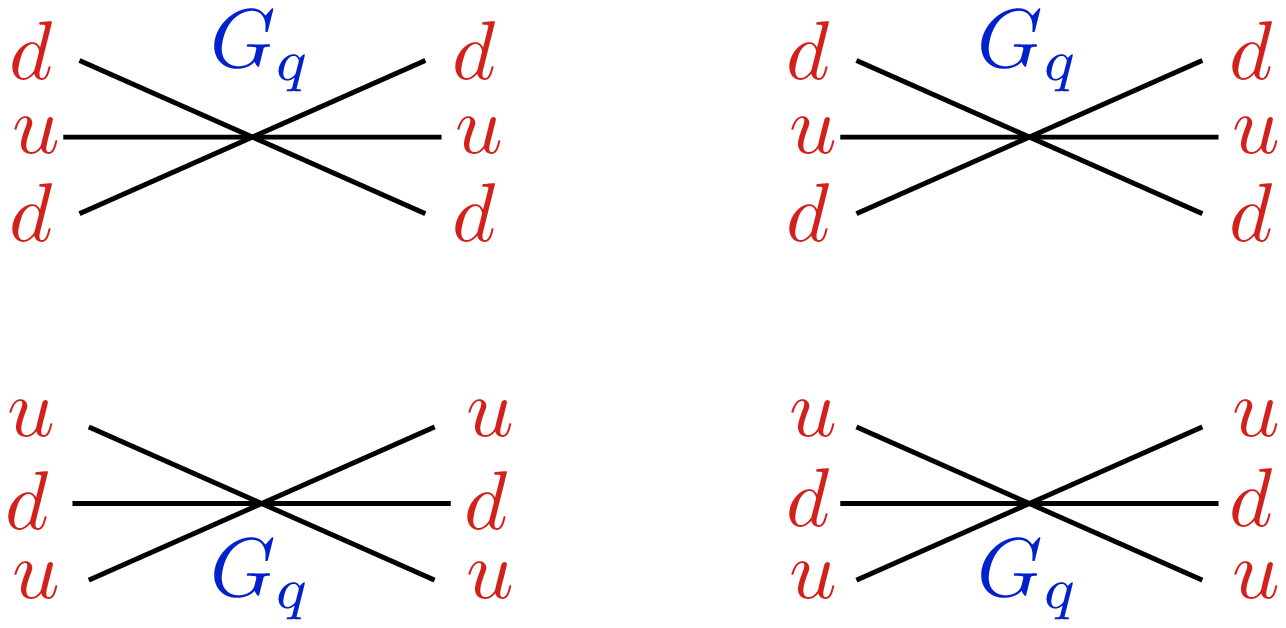}
\hspace{1cm}
\includegraphics[width=6cm ,height=5cm]{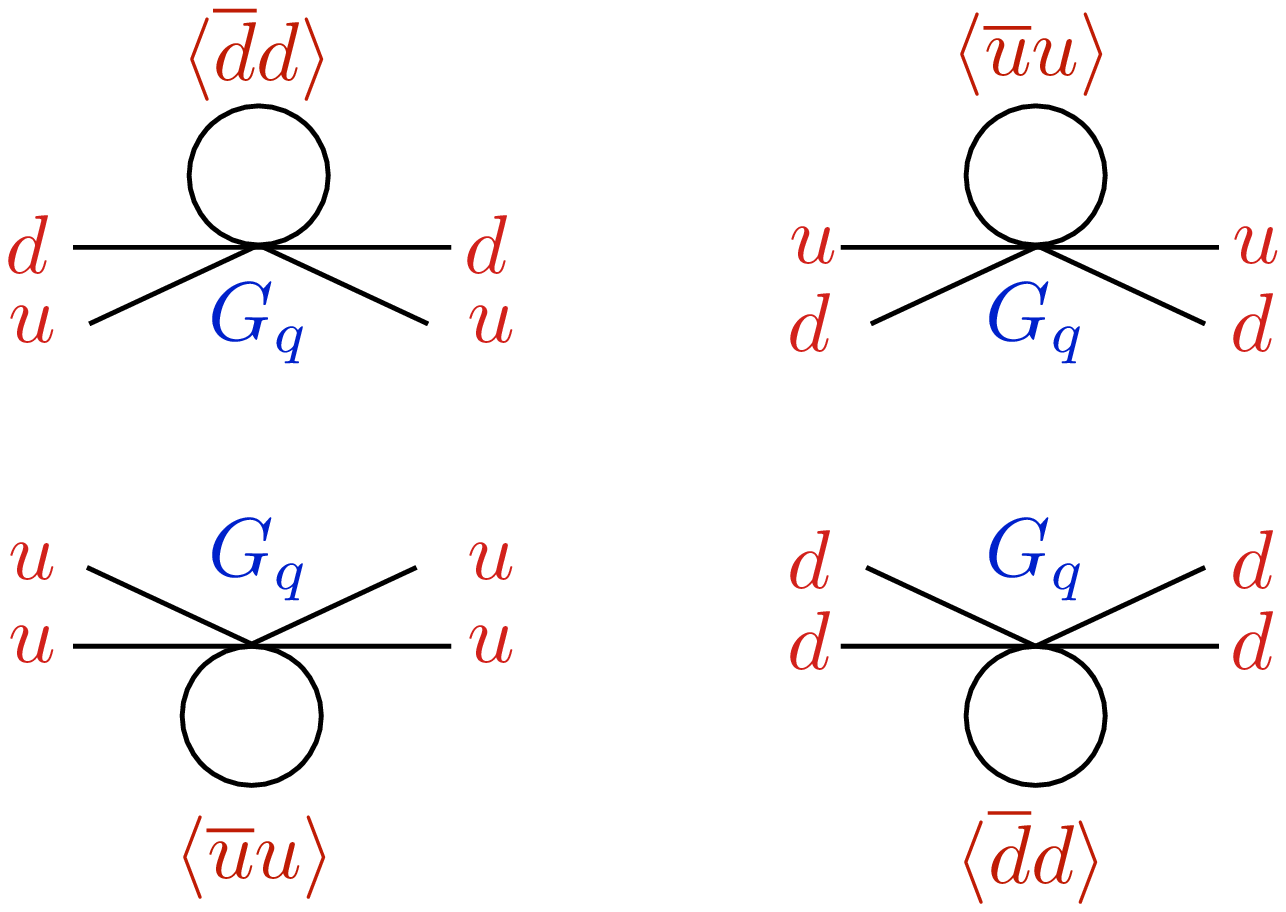}
\caption{Contact interactions generated by the mass term for the $q$-deformed 
fermion fields and their reduction from six-point to four-point by closing one 
fernion line. }
\label{Fig1}
\end{figure}

We can re-write the above equations as follows
\begin{eqnarray}
\overline{U}U &=& \left(  1+2Q~d^{\dagger}d\right)  \overline{u}u+\frac{Q^{2}}{2}~
\left(\overline{d}d\overline{u}u\overline{d}d+
\overline{d}d\overline{u}u\overline{d}d\right)  ~, \\
\overline{D}D &=& \left(  1+2Q~u^{\dagger}u\right)  \overline{d}d+\frac{Q^{2}}{2}~
\left(\overline{u}u\overline{d}d\overline{u}u+
\overline{u}u\overline{d}d\overline{u}u\right)  ~,
\end{eqnarray}
so that we identify the contact interactions between the quarks contained in
the non-interacting deformed fields Lagrangian. Figure \ref{Fig1} shows the
six-point contact interactions contained in the mass term for the $q$-deformed
quark fields.

We can reduce the six-point interactions to four-point contact terms in a mean
field approach \cite{VW}, so that we have
\begin{eqnarray}
\left(  ~\overline U U + \overline D D ~\right)   &=& \left(  1 + 2Q
\frac{\left\langle \psi^{\dagger}\psi\right\rangle }{A}\right)  \left(
~\overline u u + \overline d d ~\right) \nonumber\\
&+& \frac{Q^{2}}{2} \frac{\left\langle \overline\psi\psi\right\rangle }
{A^{2}} \left(  ~ \overline d d \overline u u + \overline d d \overline d d +
\overline u u \overline d d + \overline u u \overline u u ~\right)  ~ ,
\label{UUDD}
\end{eqnarray}
where $\langle\psi^{\dagger}\psi\rangle= \langle u^{\dagger}u\rangle= \langle
d^{\dagger}d\rangle= \rho_{\mathrm{v}}$, $\left\langle \overline\psi
\psi\right\rangle = \left\langle \overline u u \right\rangle = \left\langle
\overline d d \right\rangle = \rho_{\mathrm{s}}$, and $A=A(T;q)$ has the same
dimension of the condensate and will be determined later in this letter. The 
reduction of the six-point to four-point contact terms by closing one fermion line 
is also shown in Figure \ref{Fig1}.

Now we can write the mass term for the transformed quark fields
\begin{equation}
\mathcal{L}_{q}^{mass} =-M\overline{\Psi}\Psi=-M\left(  1+2\left\langle
\psi^{\dagger}\psi\right\rangle \Gamma\right)  ~\overline{\psi}\psi
~-\frac{M}{2}\left\langle \overline{\psi}\psi\right\rangle \Gamma
^{2}~\overline{\psi}\psi\overline{\psi}\psi~,
\end{equation}
with $\Gamma=Q/A$

\section*{Kinetic energy term}

Accordingly, the kinetic energy term for the transformed fields,
$\overline{\Psi}\gamma^{\mu}\partial_{\mu}\Psi$, can be written in terms of the standard ones as
\begin{eqnarray}
\overline{\Psi}\gamma^{\mu}\partial_{\mu}\Psi   &=& \overline{U}\gamma^{\mu}\partial_{\mu
}U+\overline{D}\gamma^{\mu}\partial_{\mu}D \nonumber \\
&=& \overline{u}\gamma^{\mu}\partial_{\mu}u+Q\left(  \overline{d}\gamma_{0}%
d\overline{u}\gamma^{\mu}\partial_{\mu}u+\overline{u}\gamma^{\mu}\partial_{\mu}u\overline
{d}\gamma_{0}d\right) \nonumber \\
&+& \overline{d}\gamma^{\mu}\partial_{\mu}d+Q\left(  \overline{u}\gamma_{0}u\overline
{d}\gamma^{\mu}\partial_{\mu}d+\overline{d}\gamma^{\mu}\partial_{\mu}%
d\overline{u}\gamma_{0}u\right)  \nonumber \\
&+& Q^{2}\left(  \overline{d}\gamma_{0}d\overline{u}\gamma^{\mu}\partial_{\mu}u\overline
{d}\gamma_{0}d\right)  +Q^{2}\left(  \overline{u}\gamma_{0}u\overline{d}\gamma^{\mu
}\partial_{\mu}d\overline{u}\gamma_{0}u\right)
\end{eqnarray}
By using an extreme mean field approximation, namely, substituting everywhere
in the kinetic energy contribution $\langle\psi^{\dagger}\psi\rangle=\langle
u^{\dagger}u\rangle=\langle d^{\dagger}d\rangle\rightarrow\rho_{\mathrm{v}}$,
and $\left\langle \overline{\psi}\psi\right\rangle =\left\langle \overline
{u}u\right\rangle =\left\langle \overline{d}d\right\rangle \rightarrow
\rho_{\mathrm{s}}$, we obtain
\begin{eqnarray}
\overline{\Psi}\gamma^{\mu}\partial_{\mu}\Psi 
&=& \overline{u}\gamma^{\mu}\partial_{\mu
}u\left(  1+2\Gamma\rho_{\mathrm{v}}\right)  \nonumber \\
&+&  \overline{d}\gamma^{\mu}\partial_{\mu}d\left(  1+2\Gamma\rho_{\mathrm{v}
}\right)  \nonumber \\
&+&  \left(  \overline{u}\gamma^{\mu}\partial_{\mu}u\right)  \Gamma^{2}
\rho_{\mathrm{v}}+\left(  \overline{d}\gamma^{\mu}\partial_{\mu}d\right)
\Gamma^{2}\rho_{\mathrm{v}} \nonumber \\
&=& \left(  \overline{u}\gamma^{\mu}\partial_{\mu}u+\overline{d}\gamma^{\mu}%
\partial_{\mu}d\right)  \left(  1+\Gamma\rho_{\mathrm{v}}\right)  ^{2}%
\end{eqnarray}
This corresponds to a usual kinetic energy with a shifted momentum
$p\rightarrow p\left(  1+\Gamma\rho_{\mathrm{v}}\right)  ^{2}$.

\section*{The full Lagrangian}

The treatment of the density dependence of the kinetic energy term is rather
cumbersome and will be postponed to a further contribution. We will consider
the influence of this momentum dependent kinetic energy term in an effective
way. Therefore, we will study a class of Lagrangians of the type%
\begin{eqnarray}
\mathcal{L}_{\mathrm{q}}^{\prime} &=& \frac{1}{\left(  1+\Gamma\rho
_{\mathrm{v}}\right)  ^{2}}\mathcal{L}_{\mathrm{q}}=\overline{\psi}\gamma^{\mu
}\partial_{\mu}\psi-M\left(  1+2\left\langle \psi^{\dagger}\psi\right\rangle
\Gamma\right)  \frac{1}{\left(  1+\Gamma\rho_{\mathrm{v}}\right)  ^{2}}
\overline{\psi}\psi\nonumber\\
&-& \frac{M}{2}\left\langle \overline{\psi}\psi\right\rangle \Gamma
^{2}\frac{1}{\left(  1+\Gamma\rho_{\mathrm{v}}\right)  ^{2}}\overline{\psi}\psi
\overline{\psi}\psi
\label{Lqprime}
\end{eqnarray}

This representative of the full Lagrangian $\mathcal{L}_{q}=\overline{\Psi}
\gamma^{\mu}\partial_{\mu}\Psi+\mathcal{L}_{q}^{mass}$, when written in terms
of the standard quark fields, can be identified with the NJL Lagrangian
\begin{equation}
\mathcal{L}_{\mathrm{NJL}}=\overline{\psi}\gamma^{\mu}\partial_{\mu}\psi -m_{0}~\overline{\psi}\psi+G~\overline{\psi}\psi
\overline{\psi}\psi~.
\label{NJL}
\end{equation}
The conditions for both Lagrangians, $\mathcal{L}_{\mathrm{NJL}}$ and
$\mathcal{L}_{\mathrm{q}}^{\prime}$, to be equivalent for any values of $T$
and $q$ are
\begin{equation}
M=\frac{\left(  1+\Gamma\rho_{\mathrm{v}}\right)  ^{2}}{\left(  1+2\Gamma
\rho_{\mathrm{v}}\right)  }m_{0}~,\label{M}%
\end{equation}
and%
\begin{equation}
G=-\frac{M}{2}\frac{\rho_{\mathrm{s}}\Gamma^{2}}{\left(  1+\Gamma
\rho_{\mathrm{v}}\right)  ^{2}}~.\label{gamma}%
\end{equation}
Inserting Eq. (\ref{M}) in Eq. (\ref{gamma}), we obtain an equation for
$\Gamma$
\begin{equation}
\Gamma^{2}-2\alpha\rho_{\mathrm{v}}~\Gamma-\alpha=0~,
\end{equation}
where
\begin{equation}
\alpha=-\frac{2G}{m_{0}\rho_{\mathrm{s}}}>0.
\end{equation}
This equation has two solutions
\begin{equation}
\Gamma_{\pm}=\alpha\rho_{\mathrm{v}}\left(  1\pm\sqrt{1+\frac{1}{\alpha
\rho_{\mathrm{v}}^{2}}}\right)  ~.\label{gpm}%
\end{equation}
The mass of the transformed fermion fields, $M$, has to be positive, so we
associate the two solutions $\Gamma_{-}$ and $\Gamma_{+}$ with the two regimes
$q<1$ and $q>1$, respectively. The quantity $A$ will be negative in both cases.

The scalar ($\rho_{\mathrm{s}}$) and vector ($\rho_{\mathrm{v}}$) densities
were calculated from the NJL model at finite temperature:
\begin{eqnarray}
\rho_{\mathrm{s}} &=& -\frac{N_{c}N_{f}}{\pi^{2}}\int_{0}^{\Lambda}
dpp^{2}\frac{m}{E}\left[  1-n-\overline{n}\right]  ~,\\
\rho_{\mathrm{v}} 
&=& \frac{N_{c}N_{f}}{\pi^{2}}\int_{0}^{\Lambda}
dpp^{2}\left[  n-\overline{n}\right]  ~,\label{rhov}
\end{eqnarray}
where
\begin{equation}
n(\mathbf{p},T,\mu)=\frac{1}{1+\exp\left[  \beta\left(  E-\mu\right)  \right]
}~,\label{nf}
\end{equation}
and
\begin{equation}
\overline{n}(\mathbf{p},T,\mu)=\frac{1}{1+\exp\left[  \beta\left(
E+\mu\right)  \right]  }~,\label{naf}%
\end{equation}
are the fermions and anti-fermions distribution functions respectively with
$E=\sqrt{\mathbf{p}^{2}+m^{2}}$.

\begin{figure}[b]
\centering
\includegraphics[width=6cm ,height=6cm]{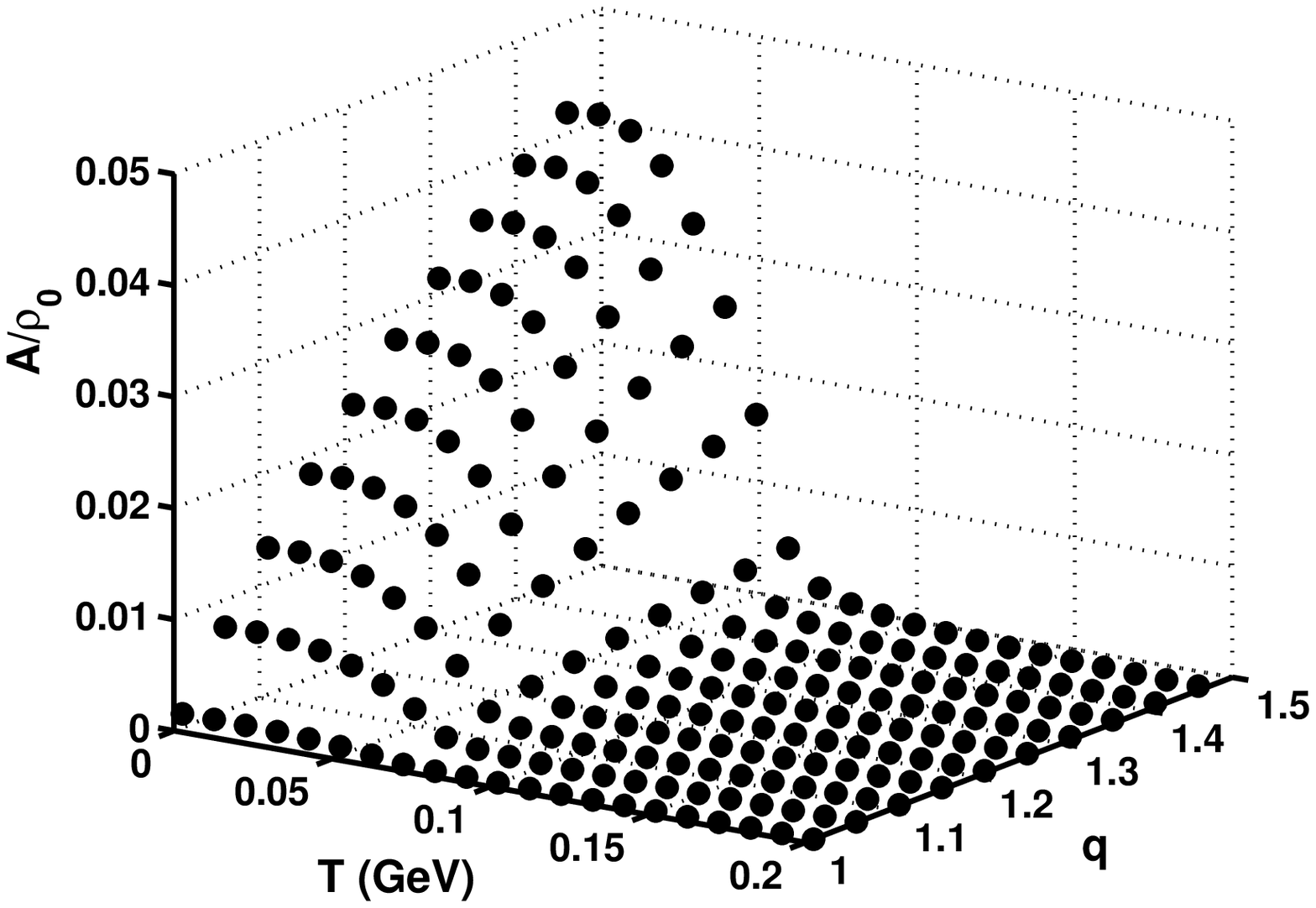}
\includegraphics[width=6cm ,height=6cm]{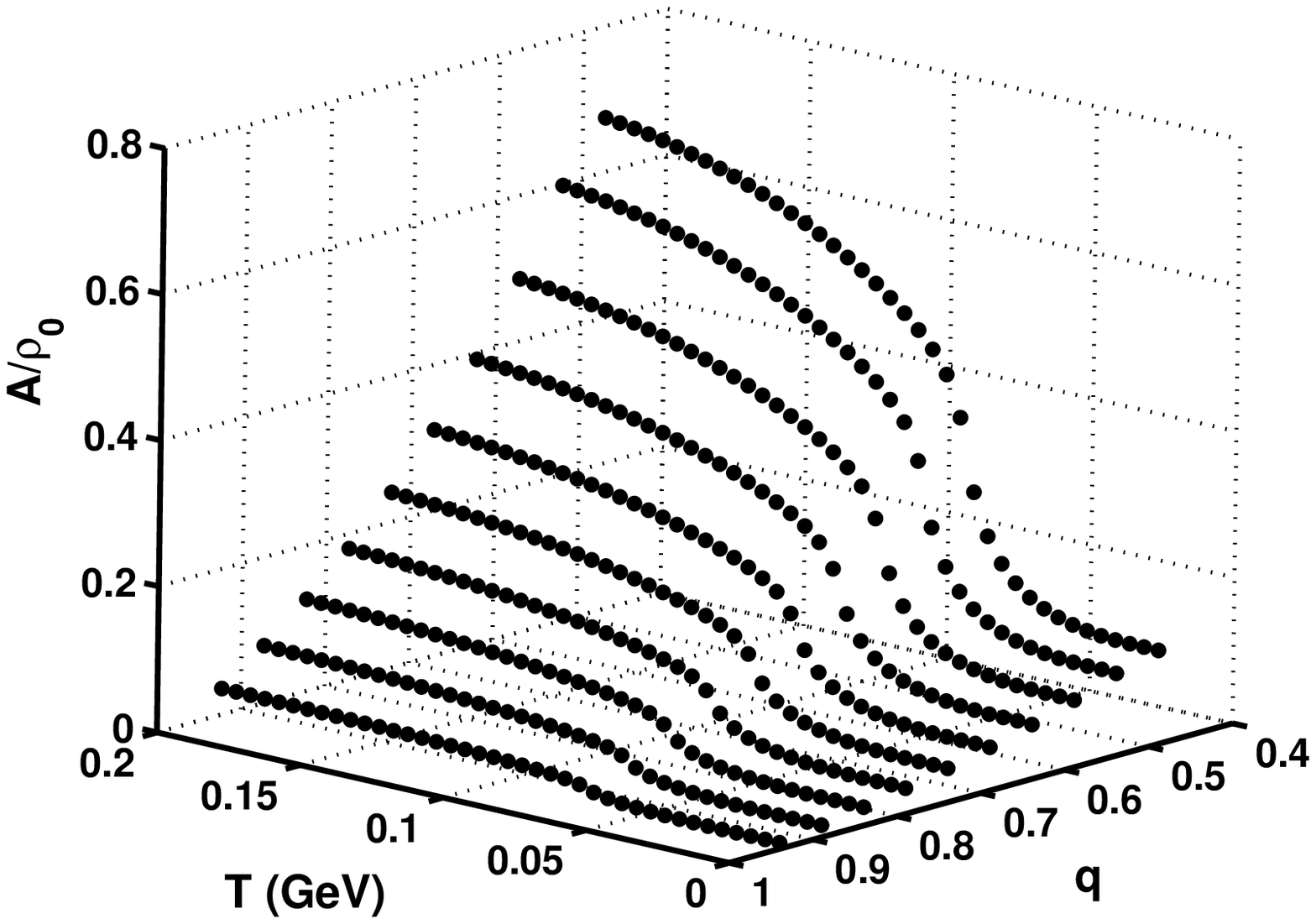}
\caption{The quantity $A$, in units of the chiral condensate at zero 
temperature $\rho_{\mathrm{s}}(T=0) = -1.42\times10^{-2}~ \mathrm{GeV}^{3}$, 
as a function of temperature and $q$-deformation for the $q>1$ and $q<1$ 
regimes.}
\label{Fig2}
\end{figure}

\begin{figure}[t]
\centering           
\includegraphics[width=10cm ,height=7cm]{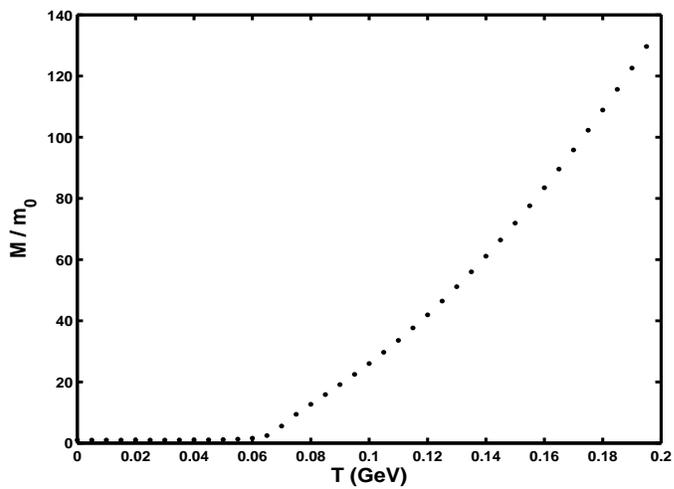}
\caption{The mass of the $q$-deformed quark fields, in units of the current 
quark mass $m_{0} = 5~\mathrm{MeV} $, as a function of $T$ for both $q>1$ and 
$q<1$ regimes. For small temperatures, $M=m_0$.}
\label{Fig3}
\end{figure}

First we solve the set of coupled gap equations for $m$, $\mu$, and
$\rho_{\mathrm{v}}$ (Eqs. \ref{mmu} and \ref{rhov}, respectively) in the NJL
model at finite temperature and chemical potential
\begin{equation}
\left\{
\begin{array}
[c]{l}%
m=m_{0}-2G\rho_{\mathrm{s}}~,\\
\mu=\mu_{0}-\frac{G}{N_{c}}\rho_{\mathrm{v}}~.
\end{array}
\right.  \label{mmu}%
\end{equation}
The next step is to calculate the scalar and vector densities entering in the
equation for $\Gamma$ for a given value of the transformation parameter $q$. In
this way we obtain $A(T;q)$, which in turn is used to obtain $M$. The
numerical results are displayed in Figures \ref{Fig2} and \ref{Fig3}, where we
show the quantity $A$ and the mass $M$ as a function of both temperature and
transformation parameter in the $q>1$ and $q<1$ regimes. It is worth to note that the mass
of the transformed fermion fields does not depend on the transformation parameter. 

The well known results of the NJL model are mapped through $A(T;q)$ from the 
non-interacting transformed fermion fields Lagrangian.
It is worth to note that the mass of the $q$-deformed fermion fields does 
not depend on the deformation of the algebra.

The quantity $A\left(  T;q\right)  $ maps the simple non-interacting model
into the NJL model. It represents, in an effective way, the correlations
introduced by the transformations, when we write the non-interacting
Lagrangian in terms of the standard quark fields. These correlations, in a
mean field approximation, are effectively represented by contact interactions
of the NJL type. It is also important to mention that it inherits the phase
transition. When the condensate and the dynamical mass vanishes with
increasing $T$, the quantity $A$ also experiences the phase transition. This
is an expected behavior, since it depends on the dynamical mass. For a given
temperature, $T$, and transformation parameter, $q$, there is a value of the mapping
function, $A(T;q)$, that makes the Lagrangians Eq.(\ref{Lqprime}) and
Eq.(\ref{NJL}) equivalent.

Summarizing, we have shown that it is possible to describe the dynamics of an
interacting system of the NJL type with a simple non-interacting system by
using a set of quantum algebra inspired transformations and a mapping function.

\vspace{0.5cm}

\textbf{Acknowledgments}

\vspace{0.5cm}

C. L. L. thanks Profs. D. Galetti and B. M. Pimentel for most helpful discussions.
This work was partially supported by FAPESP Grant No. 2002/10896-7. V.S.T.
would like to thank FAEPEX/UNICAMP for financial support.

\clearpage            


\newpage

\end{document}